# Phase-Separation Control of $K_xFe_{2-y}Se_2$ Superconductor Through Rapid-Quenching Process

Yusuke Yanagisawa[1,2]*, Masashi Tanaka[1], Aichi Yamashita[1,2], Kouji Suzuki[1,2]

Hiroshi Hara[1,2], Mohammed ElMassalami[3], Hiroyuki Takeya[1], and Yoshihiko Takano[1,2]

[1]*MANA, National Institute for Materials Science,*

*1-2-1 Sengen, Tsukuba, Ibaraki 305-0047, Japan*

[2] *Graduate School of Pure and Applied Sciences, University of Tsukuba,*

*1-1-1 Tennodai, Tsukuba, Ibaraki 305-8577, Japan*

[3]*Instituto de Fisica, Universidade Federal do Rio de Janeiro,*

*Caixa Postal 68528, 21945-970, Rio de Janeiro, Brazil*

* Corresponding Author: Yusuke Yanagisawa

E-mail: YANAGISAWA.Yusuke@nims.go.jp

Address:

Nano-Frontier Superconducting Materials Group

MANA, National Institute for Materials Science (NIMS)

1-2-1 Sengen, Tsukuba, Ibaraki 305-0047, Japan

**Abstract**

$K_xFe_{2-y}Se_2$ exhibits iron-vacancy ordering at $T_s$ ~270°C and separates into two phases: a minor superconducting (iron-vacancy-disordered) phase and a major non-superconducting (iron-vacancy-ordered) phase. The microstructural and superconducting properties of this intermixture can be tuned by an appropriate control of the quenching process through $T_s$. A faster quenching rate leads to a finer microstructure and a suppression of formation of the non-superconducting phase by up to 50%. Nevertheless, such a faster cooling rate induces a monotonic reduction in the superconducting transition temperature (from 30.7 to 26.0 K) and, simultaneously, a decrease in the iron content within the superconducting phase such that the compositional ratio changed from $K_{0.35}Fe_{1.83}Se_2$ to $K_{0.58}Fe_{1.71}Se_2$.

FeSe-based superconductors have attracted much interest because they exhibit high superconducting transition temperature ($T_c$): although $T_c$ of the parent compound FeSe is only 8 K[1], the application of high pressure (~6 GPa) was shown to enhance $T_c$ up to ~37 K.[2] It is noteworthy that intercalating potassium ions into the interstitial site between FeSe layers leads to the formation of $K_xFe_{2-y}Se_2$ with $T_c$ reaching up to 32 K at ambient pressure[3,4] and ~48 K under high pressure (~12 GPa).[5,6] Very recently, a higher $T_c$ (up to 65 K or even above 100 K) has been reported for monolayer FeSe film deposited on a $SrTiO_3$ substrate.[7,8]

The above-mentioned activities indicate that FeSe-based superconductors are of interest as the next generation of high-$T_c$ superconductors. However, it happens that the $K_xFe_{2-y}Se_2$ superconductor has one disadvantage. It contains some deficiencies in both potassium and iron sites mainly due to the conservation law of charge neutrality. Because of these deficiencies, the microstructure shows a two-phase separation: one is the iron-vacancy-disordered superconducting phase and the other is the iron-vacancy-ordered antiferromagnetic insulating phase.[9-12] Because of such a microscopic phase separation, it has been extremely difficult to understand and optimize the intrinsic properties of the superconducting phase.

Previous studies have shown that the phase separation occurs below the vacancy ordering temperature $T_s$ ~270°C.[13-15] It was also reported that the superconducting phase can be regarded as a remnant of the high temperature ($T > T_s$) phase with some compositional change.[16] Furthermore, it was suggested that the excess iron atoms that are expelled from the vacancy-ordered insulating phase will eventually aggregate within the (iron-rich) superconducting phase during the phase separation.[16] Then, a quench through $T_s$ is expected to strongly affect the microstructural and superconducting properties (see Ref. 15 and references therein), but the relationship between the compositional ratio and superconducting properties is yet unclear. Earlier, Tanaka et al.[15] followed the evolution of the superconductivity, microstructure, and compositional ratio when the quenching temperature is varied: In particular, it was demonstrated that crystals quenched above $T_s$ exhibit a mesh-like texture with the area ratio of ~34%, and $T_c$ ~31 K. On the other hand, slow-cooled crystals (or quenched below $T_s$) exhibit island-like domains with the area ratio of ~12%, and $T_c$ ~44 K. It was also noted that the island-like domain has a higher Fe content than the mesh-like one. More recently, it has been revealed that the island-like domain with $T_c$ ~44 K in the slow-cooled crystal is composed of the $K_xFe_2Se_2$ structure with perfect FeSe layers by a transmission electron microscope using the micro-sampling technique.[17]

The above studies imply that $T_c$ strongly depends on the Fe content, and provide motivation for systematic investigations on the compositional change. Further systematic studies on the effect of the cooling rate through $T_s$ will lead to a deeper understanding of the electronic phase diagram or mechanism of superconductivity in this material. In this work, we focused on the evolution of the microstructural, compositional, and superconducting properties with the variation of the quenching rate. For that purpose, we synthesized $K_xFe_{2-y}Se_2$ single crystals and subjected them to a variety of quenching processes. These quenched crystals were then studied by various techniques, namely, energy-dispersive X-ray analysis, back-scattered electron imaging by scanning electron microscope, resistivity and magnetization measurements.

Single crystals with the nominal composition of $K_{0.8}Fe_{2.0}Se_{2.0}$ were grown in a similar way to the one-step method.[18] Firstly, the starting materials of $K_2Se$ (99%) and Fe (99.9%) powders and Se (99.999%) grains were put into an alumina or graphite crucible. The crucible was sealed within a quartz tube evacuated down to $10^{-1}$ Pa or a stainless steel tube under 1.0 atm Ar gas. Secondly, these tubes were heated to 950°C for 10 h, held there for 5 h and then cooled down to 700°C at a rate of -7°C/h. The heat treatment was performed in a vertical furnace that allowed quenching from 700°C by dropping into room-temperature water or iced water. The quenching conditions were controlled by varying the type of crucible, sealing tube, cooling medium, or the exchange gas. Hereafter, the three kind of quenching conditions are labelled as follows: Quench A (alumina crucible, evacuated quartz tube, room-temperature water), Quench B (alumina crucible, evacuated quartz tube, iced water), and Quench C (graphite crucible, Ar-filled stainless-steel tube, iced water). The cooling rate of Quench C is expected to be the highest, followed by Quench B and Quench A (see the supplemental material).[19]

The microstructure of the obtained single crystals was observed on back-scattered electron (BSE) images obtained with a scanning electron microscope (SEM; JSM-6010, JEOL) operated at an acceleration voltage of 15 kV. The compositional ratio was analyzed by energy-dispersive X-ray (EDX) spectroscopy attached to the SEM equipment. The electrical resistivity was measured by a standard four-probe method using a physical property measurement system (PPMS, Quantum Design). The magnetization was measured using a superconducting quantum interference device (SQUID) magnetometer (MPMS, Quantum Design). All sample preparations were manipulated in a glove box under Ar atmosphere.

Figures 1(a) and 1(b) show the temperature dependence of in-plane resistivity of the single crystals under the three different quenching conditions. As the cooling rate of the quenching process increases, the normal state resistivity increases and becomes a semiconducting behavior with a large hump at ~150 K. On the other hand, the onset temperature $T_c^{onset}$ is almost the same for all samples: 31.5 K (Quench A), 31.1 K (Quench B), and 31.6 K (Quench C). In contrast, the zero-resistivity temperature $T_c^{zero}$ is considerably reduced: 30.7 K (Quench A), 28.6 K (Quench B) and 26.0 K (Quench C).

Figures 1(c) and 1(d) show the temperature dependence of zero-field-cooled (ZFC) magnetization of the three samples measured at $H = 10$ Oe ($H // ab$). The shielding volume fraction decreases dramatically with increasing the cooling rate, which indicates poor connectivity between the superconducting grains. The onset temperature $T_c^{mag}$ decreases monotonically as 30.1 K (Quench A), 27.2 K (Quench B), and 24.6 K (Quench C), which is a similar tendency to that of $T_c^{zero}$.

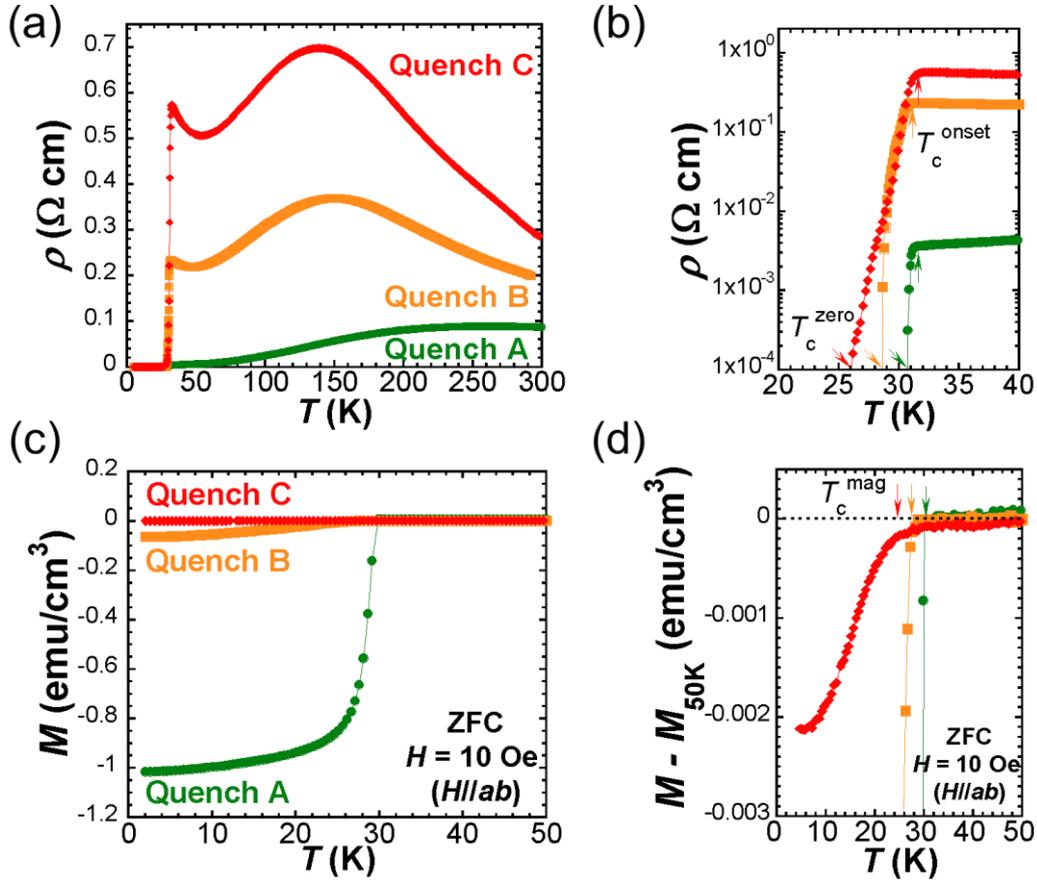

Fig. 1. (Color online) (a), (b) Temperature dependence of in-plane resistivity of the three samples under different quenching conditions: $T_c^{onset}$ and $T_c^{zero}$ are defined as the temperature at which the resistivity starts to drop and the temperature below which the resistivity becomes less than $10^{-4}$ Ω cm, respectively, as indicated by arrows in (b). (c) Temperature dependence of zero-field cooled (ZFC) magnetization of the three samples. (d) The magnetization is subtracted at a value of 50 K to see clearly the difference in diamagnetic transitions. $T_c^{mag}$ is defined as the temperature at which the diamagnetic signal starts to drop, as indicated by arrows.

The correlation between the superconducting properties and the microstructure and compositional ratio was investigated by SEM and EDX analyses. Figure 2(a) shows BSE images of the freshly cleaved surface of the three samples under different quenching conditions. All samples show clear evidence of two coexisting phases shown as a mesh-like bright texture within a dark background, in agreement with the pattern reported previously.[16,20-22] The bright and dark regions correspond to iron-rich superconducting and iron-poor insulating $K_{0.8}Fe_{1.6}Se_2$ phases, respectively. With increasing cooling rate (along the sequence Quench A to C), we see that the mesh-like network of superconducting grains becomes finer and the grain size decreases. The above-mentioned results of the shielding volume fraction may be attributed to the comparable size of superconducting grains to its penetration depth.[23] In addition, the area ratio of the bright region increases. The ratio of bright domains is estimated to be 34% (Quench A), 42% (Quench B), and 50% (Quench C) by calculation based on the binarized BSE images. A faster quenching rate leads to a suppression of the formation of the non-superconducting phase by up to 50%.

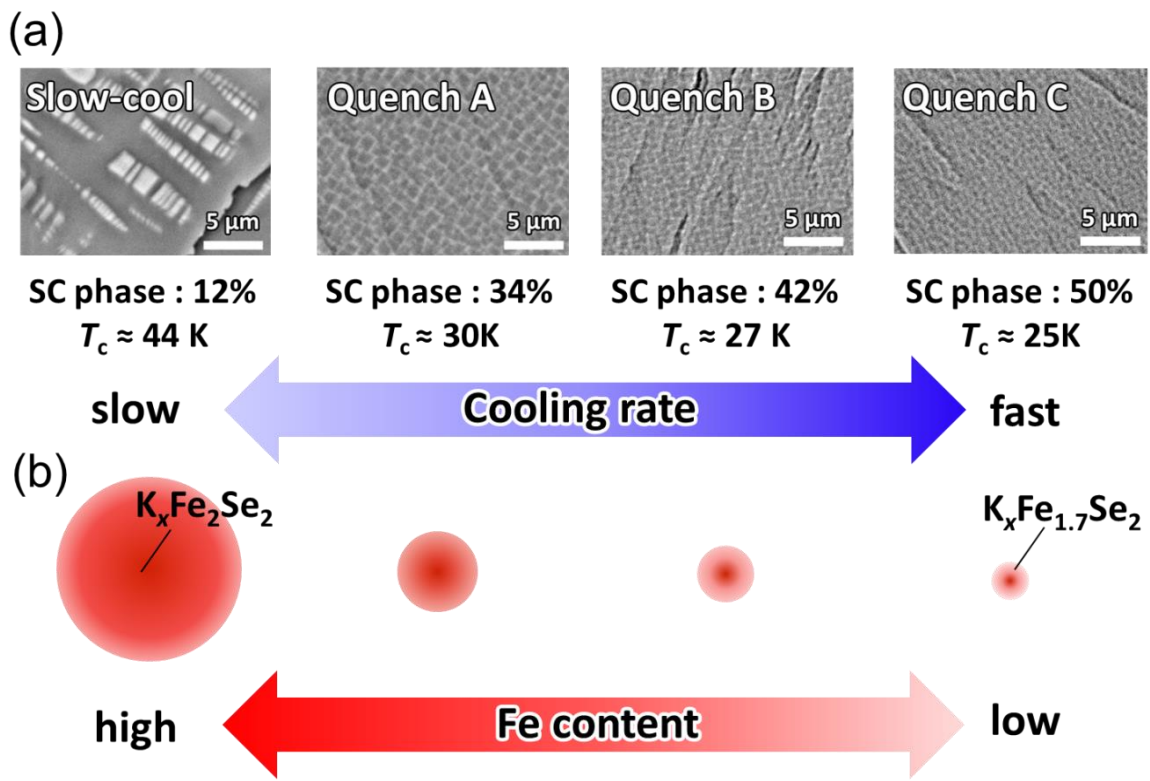

Fig.2. (Color online) (a) Back-scattered electron images of SEM measurements on cleaved surfaces of differently heat-treated samples. The slow-cooled sample is included for comparison.[15] (b) Schematic illustrations of compositional evolution of the superconducting grains and its relation to microstructures and superconducting properties, corresponding to (a).

By EDX analysis, we characterized the compositional ratio of the superconducting phase. Before discussing the obtained results, it is worth remarking that the spatial resolution of EDX spot analysis is limited to the excitation volume of the employed characteristic X-rays. Monte Carlo simulations by CASINO[24] show that the excitation volume of Fe K$\alpha$ X-ray at an acceleration voltage of 15 kV is less than 1.8 µm in depth and 1.0 µm in lateral radius. Thus, EDX spot analysis is unsuitable for resolving the difference in the compositional ratio of the submicron-sized bright and dark domains. However, EDX area analysis yields an accurate compositional ratio of the whole region (averaged bright and dark domains).

Here, we estimate the compositional ratio $K_xFe_{2-y}Se_2$ of the bright (superconducting) domain ($x_{bright}$, $y_{bright}$) using the data of the EDX area analysis ($x_{whole}$, $y_{whole}$) and the area ratio ($R_{bright}$, $R_{dark}$), and in addition, assuming the compositional ratio of the dark region to be $K_{0.8}Fe_{1.6}Se_2$ ($x_{dark} = 0.8$, $y_{dark} = 0.4$), as given by

$$\begin{cases} x_{bright} = (x_{whole} - R_{dark}x_{dark})/R_{bright} \\ y_{bright} = (y_{whole} - R_{dark}y_{dark})/R_{bright} \end{cases} \quad (1)$$

The calculated values ($x_{bright}$, $y_{bright}$) are (0.35, 0.17) in Quench A, (0.53, 0.26) in Quench B and (0.58, 0.29) in Quench C. Namely, the compositional ratios are $K_{0.35}Fe_{1.83}Se_2$ (Quench A), $K_{0.53}Fe_{1.74}Se_2$ (Quench B), and $K_{0.58}Fe_{1.71}Se_2$ (Quench C). Table I summarizes the transition temperature, compositional ratios, and area ratio of the studied samples. It is emphasized that the average iron content within the superconducting domains decreases with increasing the cooling rate along the sequence Quench A, B to C, which is best illustrated in Fig. 2(b). A faster cooling rate reduces aggregation of the iron and gives rise to a smaller superconducting grain with less iron content, and in addition, decreases the connectivity of grains. In contrast, a slower cooling rate promotes the size and iron aggregation of the superconducting grains. These findings are summarized in Fig. 3. The quenching process influences on the iron content (denoted as 2-$y$ in the superconducting $K_xFe_{2-y}Se_2$ phase) which in turn affects both $T_c$ and the area ratio. At a higher limit of iron content ($y \rightarrow 0$), the composition becomes $K_xFe_2Se_2$ and $T_c$ can be extrapolated to values in the vicinity of 44 K. Indeed, Tanaka et al.[17] reported that the superconducting $T_c$ ~44 K phase can be considered to be the iron-vacancy-free $K_xFe_2Se_2$ phase. Alternatively, Fig. 3 indicates that an increase of the iron vacancy within $K_xFe_{2-y}Se_2$ leads to a suppression of the superconductivity, and this can be understood as a decrease in iron content and the electron carriers.

Table I. Summary of the transition temperature ($T_c^{onset}$, $T_c^{zero}$ and $T_c^{mag}$), compositional ratio $K_xFe_{2-y}Se_2$ of the whole region ($x_{whole}$, $y_{whole}$; characterized by EDX area analysis), superconducting (SC) region ($x_{bright}$, $y_{bright}$; calculated), and the area ratio ($R_{bright}$) in the BSE image of the studied samples.

|  | Quench A | Quench B | Quench C |
| --- | --- | --- | --- |
| $T_c^{onset}$ (K) | 31.5 | 31.1 | 31.6 |
| $T_c^{zero}$ (K) | 30.7 | 28.6 | 26.0 |
| $T_c^{mag}$ (K) | 30.1 | 27.2 | 24.6 |
| Compositional ratio of whole region (EDX area analysis) | $K_{0.65}Fe_{1.68}Se_2$ | $K_{0.69}Fe_{1.66}Se_2$ | $K_{0.69}Fe_{1.66}Se_2$ |
| Compositional ratio of SC region (calculated) | $K_{0.35}Fe_{1.83}Se_2$ | $K_{0.53}Fe_{1.74}Se_2$ | $K_{0.58}Fe_{1.71}Se_2$ |
| Area ratio of bright region (%) | 34 | 42 | 50 |

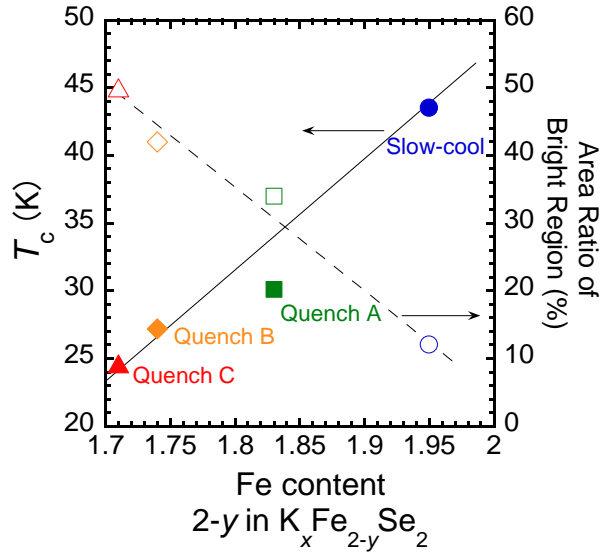

Fig.3. (Color online) Relationships among $T_c^{mag}$, area ratio of the bright region, and Fe content of all studied samples. The slow-cooled sample is included as a reference.[15] The open and closed symbols represent the area ratio and $T_c^{mag}$, respectively.

The above-described results suggest that it is difficult to obtain the pure superconducting phase because faster quenching increases the superconducting region but also reduces the iron content and leads to the suppression of superconductivity. It is expected to be necessary to increase the iron content in the solid solution and get more iron-rich superconducting region by further rapid quenching. However, this is related to limitation of solubility of the excess iron,[25] which is expected to be a small amount of iron ($1.6 < 2-y < 1.7$), as suggested in the results of the EDX area analysis (see Table I). Nevertheless, our results indicate that it is possible to optimize the microstructural and superconducting properties by appropriately tuning the quenching conditions.

In conclusion, we have shown that the microstructural, compositional and superconducting properties of $K_xFe_{2-y}Se_2$ can be controlled by the conditions of the quenching through $T_s$. In particular, $T_c$ and the ratio of the bright area in the BSE image were found to critically depend on the iron content within the superconducting grains. Our analysis indicates that a faster quenching rate does not guarantee an increase in the iron content nor the connectivity of grains. Furthermore, we have found it difficult to obtain the pure superconducting phase: this is related to limitation of solubility of the excess iron in the solid solution during crystal growth, mainly due to the conservation law of charge neutrality. Nevertheless, it is expected to be possible to optimize the superconducting properties, in particular, the connectivity of the iron-rich superconducting grains by appropriate tuning of the quenching conditions.

**Acknowledgment**

This work was partially supported by CREST, Japan Science and Technology Agency (JST), and KAKENHI (Grants-in-Aid for Scientific Research, Grant No. 25420697), Japan Society for the Promotion of Science (JSPS).

Corresponding Author's E-mail: YANAGISAWA.Yusuke@nims.go.jp

**Reference List**

1) F. C. Hsu, J. Y. Luo, K. W. Yeh, T. K. Chen, T. W. Huang, P. M. Wu, Y. C. Lee, Y. L. Huang, Y. Y. Chu, D. C. Yan, and M. K. Wu, Proc. Natl. Acad. Sci. **105**, 14262 (2008).

2) S. Margadonna, Y. Takabayashi, Y. Ohishi, Y. Mizuguchi, Y. Takano, T. Kagayama, T. Nakagawa, M. Takata, and K. Prassides, Phys. Rev. B **80**, 064506 (2009).

3) J. G. Guo, S. F. Jin, G. Wang, S. C. Wang, K. X. Zhu, T. T. Zhou, M. He, and X. L. Chen, Phys. Rev. B **82**, 180520(R) (2010).

4) Y. Mizuguchi, H. Takeya, Y. Kawasaki, T. Ozaki, S. Tsuda, T. Yamaguchi, and Y. Takano, Appl. Phys. Lett. **98**, 042511 (2011).

5) L. Sun, X. J. Chen, J. Guo, P. Gao, Q. Z. Huang, H. Wang, M. Fang, X. Chen, G. Chen, Q. Wu, C. Zhang, D. Gu, X. Dong, L. Wang, K. Yang, A. Li, X. Dai, H. K. Mao, and Z. Zhao, Nature **483**, 67 (2012).

6) Y. Yamamoto, H. Yamaoka, M. Tanaka, H. Okazaki, T. Ozaki, Y. Takano, J. F. Lin, H. Fujita, T. Kagayama, K. Shimizu, N. Hiraoka, H. Ishii, Y. F. Liao, K. D. Tsuei, and J. Mizuki, Sci. Rep. **6**, 30946 (2016).

7) S. He, J. He, W. Zhang, L. Zhao, D. Liu, X. Liu, D. Mou, Y. B. Ou, Q. Y. Wang, Z. Li, L. Wang, Y. Peng, Y. Liu, C. Chen, L. Yu, G. Liu, X. Dong, J. Zhang, C. Chen, Z. Xu, X. Chen, X. Ma, Q. Xue, and X. J. Zhou, Nat. Mater. **12,** 605 (2013).

8) J. F. Ge, Z. L. Liu, C. Liu, C. L. Gao, D. Qian, Q. K. Xue, Y. Liu, and J. F. Jia, Nat. Mater. **14**, 285 (2015).

9) H. H. Wen, Rep. Prog. Phys. **75**, 112501 (2012).

10) E. Dagotto, Rev. Mod. Phys. **85**, 849 (2013).

11) W. Bao, J. Phys.: Condens. Matter **27**, 023201 (2015).

12) A. Krzton-Maziopa, V. Svitlyk, E. Pomjakushina, R. Puzniak, and K. Conder, J. Phys.: Condens. Matter **28**, 293002 (2016).

13) W. Bao, Q. Z. Huang, G. F. Chen, M. A. Green, D. M. Wang, J. B. He, and Y. M. Qiu, Chin. Phys. Lett. **28**, 086104 (2011).

14) A. Ricci, N. Poccia, B. Joseph, G. Arrighetti, L. Barba, J. Plaisier, G. Campi, Y. Mizuguchi, H. Takeya, Y. Takano, N. L. Saini, and A. Bianconi, Supercond. Sci. Technol. **24**, 082002 (2011).

15) M. Tanaka, Y. Yanagisawa, S. J. Denholme, M. Fujioka, S. Funahashi, Y. Matsushita, N. Ishizawa, T. Yamaguchi, H. Takeya, and Y. Takano, J. Phys. Soc. Jpn. **85**, 044710 (2016).

16) Y. Liu, Q. Xing, W. E. Straszheim, J. Marshman, P. Pedersen, R. McLaughlin, and T. A. Lograsso, Phys. Rev. B **93**, 064509 (2016).

17) M. Tanaka, H. Takeya, and Y. Takano, Appl. Phys. Express **10**, 23101 (2017).

18) T. Ozaki, H. Takeya, H. Okazaki, K. Deguchi, S. Demura, Y. Kawasaki, H. Hara, T. Watanabe, T. Yamaguchi, and Y. Takano, Europhys. Lett. **98**, 27002 (2012).

19) (Supplemental material) The details of the quenching conditions and consideration of the cooling rate are provided online.

20) Y. Liu, Q. Xing, K. W. Dennis, R. W. McCallum, and T. A. Lograsso, Phys. Rev. B **86**, 144507 (2012).

21) X. Ding, D. Fang, Z. Wang, H. Yang, J. Liu, Q. Deng, G. Ma, C. Meng, Y. Hu, and H. H. Wen, Nat. Commun. **4**, 1897 (2013).

22) Z. Wang, Y. Cai, Z. W. Wang, C. Ma, Z. Chen, H. X. Yang, H. F. Tian, and J. Q. Li, Phys. Rev. B **91**, 064513 (2015).

23) D. A. Torchetti, M. Fu, D. C. Christensen, K. J. Nelson, T. Imai, H. C. Lei, and C. Petrovic, Phys. Rev. B

**83**, 104508 (2011).

24) D. Drouin, A. R. Couture, D. Joly, X. Taster, V. Aimez, and R. Gauvin, Scanning **29**, 92 (2011).

25) D. P. Shoemaker, D. Y. Chung, H. Claus, M. C. Francisco, S. Avci, A. Llobet, and M. G. Kanatzidis, Phys. Rev. B **86**, 184511 (2012).

**Supplemental Material for:**

# Phase-Separation Control of $K_x Fe_{2-y} Se_2$ Superconductor Through Rapid-Quenching Process

Yusuke Yanagisawa[1,2]*, Masashi Tanaka[1], Aichi Yamashita[1,2], Kouji Suzuki[1,2]

Hiroshi Hara[1,2], Mohammed ElMassalami[3], Hiroyuki Takeya[1], and Yoshihiko Takano[1,2]

[1]*MANA, National Institute for Materials Science,*

*1-2-1 Sengen, Tsukuba, Ibaraki 305-0047, Japan*

[2] *Graduate School of Pure and Applied Sciences, University of Tsukuba,*

*1-1-1 Tennodai, Tsukuba, Ibaraki 305-8577, Japan*

[3]*Instituto de Fisica, Universidade Federal do Rio de Janeiro,*

*Caixa Postal 68528, 21945-970, Rio de Janeiro, Brazil*

\* Corresponding Author: Yusuke Yanagisawa

E-mail: YANAGISAWA.Yusuke@nims.go.jp

# How to Evaluate the Cooling Rate of Quenching Process

In this study, we performed three different quenching conditions by choosing the components (the crucible, sealing container and exchanger gas) and the cooling medium as described in Table S·I on the basis of Newton's law of cooling and Fourier's law of heat conduction.

According to Newton's law of cooling[1], the rate of heat transfer is proportional to the difference between the temperature of object surface and that of cooling medium. This means that the cooling rate of iced-water-quenching is higher than that of water-quenching. In addition, the ice also serves to prevent the vapor bubbling, which generally occurs on the quenched surface and acts as a thermal insulator. Therefore, iced-water-quenching is a better rapid coolant and the cooling rate of Quench B is expected to be much higher than that of Quench A.

In heat transfer, the thermal diffusivity $\alpha = k/dc_\mathrm{p}$ (m$^2$/s) is an important property: here $k$ is thermal conductivity [W/(m·K)], $d$ is the density (kg/m$^3$) and $c_\mathrm{p}$ is the specific heat under constant pressure [J/(kg·K)]. According to Fourier's law of heat conduction,[1] the cooling rate is proportional to the thermal diffusivity $\alpha$. Table S·II summarizes the thermophysical parameters of the crucible and sealing container.[2-12] As an illustration, for the graphite crucible, $\alpha$ is 97.8 mm$^2$/s (~8 times higher than that of alumina which is 2.1 mm$^2$/s). For the stainless-steel sealing tube, $\alpha$ is 3.8 mm$^2$/s (~5 times higher than that of quartz which is 0.7 mm$^2$/s). Note that, in order to change effectively the cooling rate from Quench B to Quench C, we changed the type of crucible and sealing container but also reduced the volume (see Table S·I). Therefore, the cooling rate of Quench C is expected to be the highest, followed by Quench B and Quench A.

**Table S·I.** Quenching conditions of the samples: the dimension of crucible and sealing container is described as the inner diameter (ID) x outer diameter (OD) x length (L).

| Sample No. | Sealing container ID x OD x L [mm] | Crucible ID x OD x L [mm] | Cooling medium |
|---|---|---|---|
| Quench A | evacuated quartz tube 9 x 11 x 60 | alumina 6.5 x 8.5 x 20 | water |
| Quench B | evacuated quartz tube 9 x 11 x 60 | alumina 6.5 x 8.5 x 20 | iced water |
| Quench C | Ar-filled stainless-steel tube 7.5 x 9.5 x 60 | graphite 6 x 7.4 x 20 | iced water |

**Table S·II.** Thermal conductivity $k$,[2-5] heat capacity at constant pressure $c_p$,[5-8] density $d$ [9-12] and thermal diffusivity $\alpha$ of the crucible and sealing container.

|  | $k$ [W/(m·K)] | $c_p$ [J/(kg·K)] | $d$ [kg/m$^3$] | $\alpha$ [mm$^2$/s] |
|---|---|---|---|---|
| Alumina | 36.5 | 766 | 3940 | 12.1 |
| Graphite | 128.0 | 744 | 1760 | 97.8 |
| Quartz | 1.2 | 812 | 2200 | 0.7 |
| Stainless steel (SUS304) | 14.9 | 499 | 7870 | 3.8 |

References

1) F. P. Incropera, D. P. DeWitt, T. Bergman, and A. Lavine, *Introduction to Heat Transfer* (John Wiley & Sons, New York, 2011) 6th ed., Chap. 1-2.
2) St. Burghartz and B. Schulz, J. Nucl. Mater. **212-215**, 1065 (1994).
3) C. Y. Ho, R. W. Powell, and P. E. Liley, J. Phys. Chem. Ref. Data **1**, 279 (1972).
4) K. L. Wrat and T. J. Connolly, J. Appl. Phys. **30**, 1702 (1959).
5) R. H. Bogaard, P. Desai, H. H. Li, and C. Y. Ho, Thermochim. Acta. **218**, 373 (1993).
6) D. C. Ginnings and R. J. Corruccini, J. Res. Natl. Bur. Stand. **38**, 593 (1947).
7) W. DeSorbo and W. W. Tyler, J. Chem. Phys. **26**, 244 (1957).
8) B. S. Hemingway, Am. Min. **72**, 273 (1987).
9) R. L. Coble and W. D. Kingery, J. Am. Ceram. Soc. **39**, 377 (1956).
10) M. Akoshima and T. Baba, Int. J. Thermophys. **26**, 151 (2005).
11) H. J. McSkimin, J. Appl. Phys. **24**, 988 (1953).
12) R. S. Graves, T. G. Kollie, D. L. McElroy, and K. E. Gilchrist, Int. J. Thermophys. **12**, 409 (1991).